# A Hybrid Sampling and Multi-Objective Optimization Approach for Enhanced Software Defect Prediction

Jie Zhang
School of Computer Science
China University of Geosciences
Wuhan, China

Dongcheng Li*
Department of Computer Science
California State Polytechnic
University - Humboldt,
Arcata, USA

W. Eric Wong
Department of Computer Science
University of Texas at Dallas
Richardson, USA

Shengrong Wang
School of Computer Science
China University of Geosciences
Wuhan, China

*Abstract*—Accurate early prediction of software defects is essential to maintain software quality and reduce maintenance costs. However, the field of software defect prediction (SDP) faces challenges such as class imbalances, high-dimensional feature spaces, and suboptimal prediction accuracy. To mitigate these challenges, this paper introduces a novel SDP framework that integrates hybrid sampling techniques—specifically Borderline SMOTE and Tomek Links—with a suite of multi-objective optimization algorithms, including NSGA-II, MOPSO, and MODE. The proposed model applies feature fusion through multi-objective optimization, enhancing both the generalization capability and stability of the predictions. Furthermore, the integration of parallel processing for these optimization algorithms significantly boosts the computational efficiency of the model. Comprehensive experiments conducted on datasets from NASA and PROMISE repositories demonstrate that the proposed hybrid sampling and multi-objective optimization approach improves data balance, eliminates redundant features, and enhances prediction accuracy. The experimental results also highlight the robustness of the feature fusion approach, confirming its superiority over existing state-of-the-art techniques in terms of predictive performance and applicability across diverse datasets.

## I. INTRODUCTION

Software systems have been applied in various fields such as banking, healthcare, and flight control. The various defects present in software products can pose serious threats to operational efficiency and even to life and property safety [1]. However, software defects are inevitable. Software defect prediction technology (SDP) offers a solution for early detection of defects during software development, helping to reduce testing and maintenance costs and improve software quality.

Software defect prediction involves constructing models based on information from software repositories before conducting software testing. The goal is to predict the distribution of defects in the program modules to be tested, which provides a basis for allocating software testing resources [2]. Identifying these software defects in the early stages of software engineering can reduce testing costs [3]. Current research in software defect prediction faces challenges such as class imbalance in data, high feature dimensionality, and low predictive accuracy.

The main contributions of this paper are as follows:

- This paper develops a software defect prediction model, with a focus on enhanced data processing and feature selection.

- To address the issue of class imbalance in data in SDP data processing, this paper employs a hybrid sampling algorithm based on Borderline SMOTE and Tomek Links. By combining the advantages of over-sampling and under-sampling algorithms, the model increases defective samples while addressing the issue of decision boundary overlap caused by new synthetic samples. This achieves a global balance between defective and non-defective samples in the dataset, thus enhancing the model's predictive performance and stability.

- To address the high-dimensionality problem in SDP feature selection, this paper transforms feature selection into a multi-objective optimization problem. It employs NSGA-II, MOPSO, and MODE algorithms to identify an effective subset of features, enhancing the model's predictive validity and accuracy.

- To enhance feature selection efficiency in software defect prediction, reducing iterations in multi-objective optimization is crucial. However, this can widen the gap between the Pareto front and the true optimal solution, hindering algorithm convergence. This paper proposes a feature fusion scheme based on voting and weighting methods. This paper introduces a feature fusion scheme that combines voting and weighting methods to mitigate this issue. By balancing the feature selection preferences of optimization algorithms like MOPSO, NSGA-II, and MODE, the proposed approach improves the classification accuracy of software defect prediction models.

The remainder of this paper is organized as follows. Section 2 reviews the current state of related research. Section 3 introduces the software defect prediction model developed in this study. Section 4 describes the experiments conducted to evaluate the applicability and effectiveness of the proposed model based on the result and analysis. Finally, the paper concludes with a summary of the work presented.

## II. RELATED STUDIES

This section reviews related research in software defect prediction, focusing on two main areas: prediction based on imbalanced data and prediction using multi-objective optimization algorithms.

### A. Software Defect Prediction with Imbalanced Data

Researchers have proposed several methods for handling imbalanced data in software defect prediction. Chawla et al. [4] proposed an oversampling method called the Synthetic Minority Oversampling Technique (SMOTE). SMOTE synthesizes samples from the minority class to balance the dataset. However, it has issues with data marginalization and overlapping. To address these, Han et al. [5] proposed Borderline-SMOTE, which focuses on generating new samples near the decision boundary to avoid marginalization.

Wilson [6] addressed the issue of overlapping samples in data by proposing the Edited Nearest Neighbor (ENN) algorithm. This algorithm reduces overlapping samples by removing those samples from the dataset that differ from the target sample in the majority of their nearest neighbors. ENN reduces the number of overlapping samples, thereby enhancing the quality of the dataset for classification tasks.

He et al. [7] found that SMOTE could expand the minority sample region. This expansion can lead to misclassification. They proposed the Adaptive Synthetic Sampling (ADASYN). This algorithm analyzes the distribution of minority class samples in the neighborhood and generates new samples around these minority samples, reducing the misclassification rate of majority class samples. Bennin et al. [8] introduced the MAHAKIL algorithm, which uses genetic algorithms to iteratively generate new samples.

Most of the above methods achieve class balance in the dataset by increasing minority class samples. Some researchers have added constraints to control the selection process to reduce the loss of majority class information. Mani and Zhang [9] proposed the Near-Miss algorithm, which selects samples based on the minimum average distance to the nearest samples but has computational limitations. Zeng et al. [10] proposed the SMOTE combined with Tomek links algorithm (SMOTE-Tomek), combining the advantages of both methods to increase minority class samples while reducing majority class samples, improving classification performance. Li et al. [11] proposed the SMPSO-HS-ASABoost algorithm to address data imbalance and feature redundancy in software defect prediction, improving prediction efficiency and accuracy.

### B. Software Defect Prediction Based on Multi-objective Optimization Algorithms

Software modules not only face data imbalance but also feature redundancy. Research shows that with an increase in redundant features, the number of training samples increases exponentially [12]. Therefore, removing redundant features can reduce computation, improve classification accuracy, and optimize classification models. In recent years, many feature selection methods based on heuristic algorithms have been proposed, such as Genetic Algorithm (GA) [13], Ant Colony Optimization (ACO) [14], Artificial Bee Colony (ABC) [15], and Particle Swarm Optimization (PSO) [16].

Bejjanki et al. [17] implemented ACO on eight different open-source datasets and compared it with Logistic Regression, K-Nearest Neighbors, and SVM algorithms. The results showed that ACO performed better than other prediction methods. Wu et al. [18] proposed the Multi-Objective Bat Algorithm (MOBA), using false positive rate and detection probability as two objective functions. Simulation results indicated that MOBA could save resource consumption and improve software quality compared to other commonly used algorithms. Gu et al. [19] used the Competitive Swarm Optimizer (CSO), a variant of PSO, to address the high-dimensional feature selection problem. Tran et al. [20] proposed a variable-length particle swarm optimization algorithm that defines a smaller search space by allowing particles to have different lengths, leading to a smaller feature subset and improving classification performance in a shorter time.

Existing imbalanced data handling methods have limited effectiveness in improving classification results, and feature selection methods struggle with efficiently selecting high-dimensional feature subsets. While multi-objective optimization algorithms have achieved some success in feature selection and classifier parameter tuning, their overall effectiveness remains limited.

## III. SOFTWARE DEFECT PREDICTION MODEL

This section addresses the issues of class imbalance in datasets and high-dimensional feature selection to develop a software defect prediction model based on hybrid sampling and multi-objective optimization algorithms. Additionally, by combining NSGA-II, MOPSO, and MODE, the Pareto front feature subsets obtained from these algorithms are optimized using feature fusion methods to enhance the model's predictive performance and applicability. The model consists of three parts (as shown in Fig. 1): data processing, model construction, and evaluation metrics.

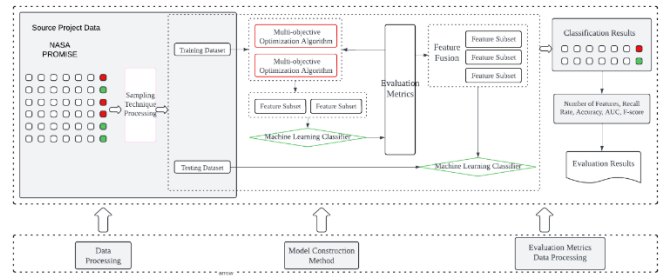

Fig. 1. Software defect prediction model.

- **Data Processing**: The datasets used in this study are NASA and PROMISE [21]. After normalizing the datasets, sampling techniques are employed to balance the data. To verify the model's reliability, the K-fold cross-validation method is used to divide the data into training and testing datasets.

- **Model Construction**: Multiple multi-objective optimization algorithms are used simultaneously with the training dataset for model training. In each iteration, machine learning classifiers use the feature subsets selected by the multi-objective optimization algorithms for defect prediction. After the specified number of iterations, each multi-objective optimization algorithm obtains feature subsets, which are then subjected to feature fusion. The trained model is then used to predict the test dataset.

- **Evaluation Metrics**: The evaluation metrics obtained from the model predictions on the test dataset.

This study presenting the basic assumptions and problem statements as follows:

- The data used in the experiments are from publicly available software defect datasets, NASA and PROMISE. To reduce the complexity of problem analysis, noise caused by manually entered incorrect labels is not considered.

- Each dataset in this study contains labeled samples, categorized as either defective or non-defective, with no unlabeled data included. To clearly define the problem boundaries, the datasets are limited to these two classes, reducing the software defect prediction task to a straightforward binary classification problem.

- The class imbalance issue in software defect datasets is categorized into inter-class imbalance and intra-class imbalance. To reduce the complexity of data sampling algorithms, only inter-class imbalance is considered, excluding intra-class imbalance.

- In transforming the feature selection problem into a multi-objective optimization problem, for ease of encoding, the weight of each feature is not considered, assuming all features have equal weight. After standardizing feature weights, a binary encoding method is used, where '0' indicates the feature is not selected, and '1' indicates the feature is selected.

- For identical feature subsets, classifiers of the same type are assumed to produce consistent and stable classification results. To simplify model design, it is assumed that classifier performance is stable, thus eliminating the need to balance the classification tendencies of different classifiers. The parameters used by a classifier are the same across different datasets.

*A. Feature Selection Based on Multi-objective Optimization*

This study uses binary encoding scheme for various multi-objective optimization algorithms. This scheme maps the solution space of the feature selection problem into the bit-string space, where genetic operations are performed. The chromosome length equals the number of features in the target dataset. For example, let $Pop = \{f_1, f_2, f_3, f_4, f_5, f_6, f_7\}$, with the initial population $\{1001100, 0101001, 0111010, ...\}$, representing feature subsets $\{f_1, f_4, f_5\}, \{f_2, f_4, f_7\}, \{f_2, f_3, f_4, f_6\}$.

After designing chromosome encoding, corresponding crossover and mutation operators are required. The crossover operator used is single-point crossover. Since chromosomes use binary encoding, the new chromosomes generated should have equal means and a ratio of approximately 1 for the difference in real values after decoding. Assuming the real values corresponding to the parent chromosomes are $p_1$ and $p_2$, the real values of the offspring after crossover and mutation are $c_1$ and $c_2$. The equation holds:

$$\frac{p_1 + p_2}{2} = \frac{c_1 + c_2}{2} \tag{1}$$

The ratio of the distance between offspring and parent chromosomes, defined as the spread factor $\beta$, is given by the formula:

$$\beta = \left|\frac{c_1 - c_2}{p_1 - p_2}\right| \tag{2}$$

The spread factor $\beta$ can be greater than, less than, or equal to 1, indicating the similarity between offspring and parents. According to existing research, when using single-point crossover with binary encoding, the offspring have equal mean values and a spread factor of 1 after decoding. Additionally, Hamming cliffs, fixed precision, and bounded variables must be considered.

The offspring chromosomes $c_1$ and $c_2$ can be calculated from the known parent chromosomes $p_1$ and $p_2$:

$$c_1 = 0.5(p_1 + p_2) - 0.5\beta(p_2 - p_1) \tag{3}$$

$$c_2 = 0.5(p_1 + p_2) + 0.5\beta(p_2 - p_1) \tag{4}$$

Once the spread factor $\beta$ is determined, new offspring can be calculated and encoded as corresponding binary chromosomes. A probability density function is used to fit the spread factor $\beta$, aiming to approximate the binary distribution as closely as possible. Following a similar probability distribution, the probability density function $c(\beta)$ [22] is presented below.

$$c(\beta) = \begin{cases} 0.5(n+1)\beta^n, & \beta \leq 1 \\ 0.5(n+1)\dfrac{1}{\beta^n}, & \beta > 1 \end{cases} \tag{5}$$

where $n$ represents the distribution index, with larger values indicating that the real values of the offspring chromosomes are closer to those of the parents. The propagation factor $\beta$, derived from the integral area of the probability density function $c(\beta)$, is given by:

$$\beta = \begin{cases} (2u)^{\frac{1}{n+1}}, & \text{if } u \leq 0.5 \\ \left(\dfrac{1}{2-2u}\right)^{\frac{1}{n+1}}, & \text{if } u > 0.5 \end{cases} \tag{6}$$

where $u$ represents the probability, with values ranging between 0 and 1. $n$ referred to as the crossover distribution index, is an arbitrary non-negative real number set by the user. The larger the value of $n$, the higher the probability that the offspring will be closer to the parents. Conversely, the smaller the value of $n$, the greater the probability that the offspring will be farther away from the parents.

The optimization objectives are to minimize the number of features while maximizing the performance of the software defect prediction model, measured by the Area Under the ROC Curve (AUC). For the feature minimization objective, denoted as min $g(x)$, the transformation function T can be defined as:

$$f(x) = \begin{cases} C_{max} - g(x), & if\, g(x) < C_{max} \\ 0, & if\, g(x) \geq C_{max} \end{cases} \quad (7)$$

where $C_{max}$ represents the total number of features in the dataset. $C_{max}$ is not a fixed value but varies depending on the dataset. For the maximization objective of increasing AUC, denoted as $max\, g(x)$, the transformation function T can be defined as:

$$f(x) = \begin{cases} g(x) - C_{min}, & if\, g(x) > C_{min} \\ 0, & if\, g(x) \leq C_{min} \end{cases} \quad (8)$$

where $C_{min}$ can be a specific input value, a theoretical minimum, or the minimum value of $g(x)$ in all or the last k generations. In this study, the AUC value ranges from 0 to 1, $C_{min}$ is defaulted to 0 without adjusting $g(x)$ results.

As shown in Fig. 2, the feature selection and fusion process in the algorithm taking NSGA-II as an example.

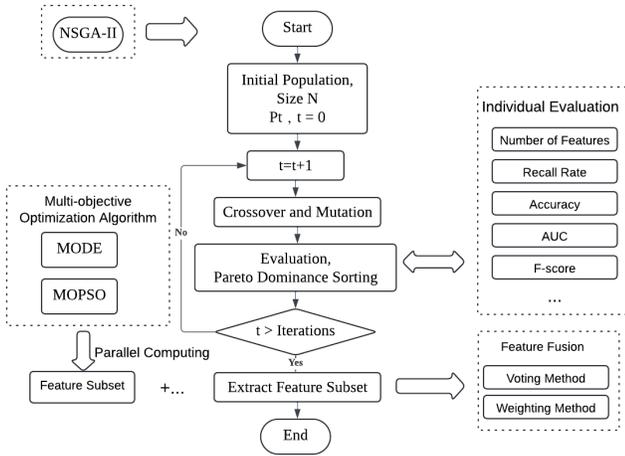

Fig. 2. Algorithm flowchart

### B. Feature Fusion Based on Multiple Multi-objective Optimization

This study proposes a feature fusion method aligned with the encoding strategy to mitigate the selection bias of multi-objective optimization algorithms. Additionally, parallel computation of multi-objective optimization algorithms is implemented to enhance the model's operational efficiency. This study introduces parallel computation to combine the NSGA-II, MODE, and MOPSO. As shown in Fig. 3, each algorithm is locked when it starts running, with the main thread lock count Flag set to 3. Once an algorithm completes, it releases the lock and decreases Flag by 1. If Flag is not zero, indicating not all algorithms have finished, the main thread remains blocked, waiting. Each algorithm separately derives feature subsets, followed by feature fusion methods to produce new ordered feature subset sequences, enhancing the model's applicability and predictive accuracy.

The feature fusion process optimizes and ranks the features in the Pareto front feature subsets, obtaining a feature sequence sorted by their contribution to the model's classification performance. To balance the selection tendencies of different intelligent optimization algorithms, this study uses feature fusion methods, including voting and weighting, for the Pareto front feature subsets. The process flow for feature fusion using voting and weighting is illustrated in Fig. 4.

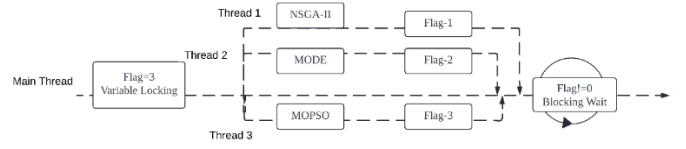

Fig. 3. Parallel computation of multi-objective optimization algorithms

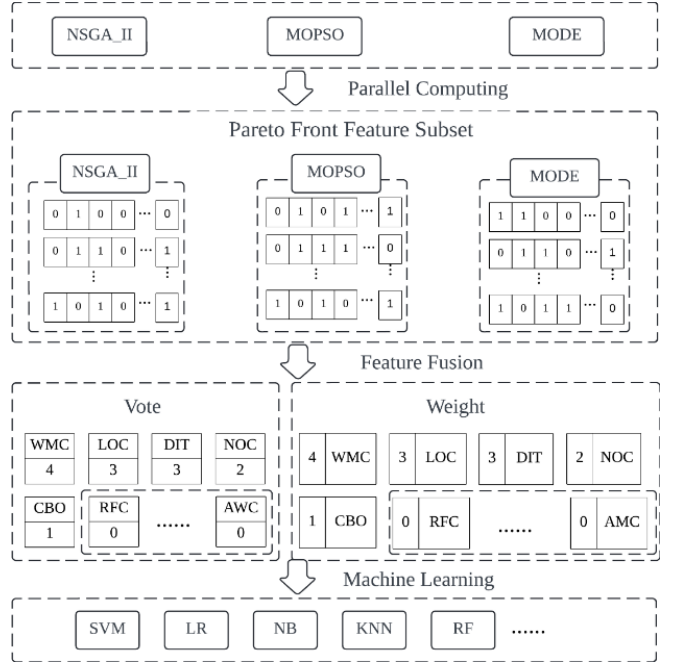

Fig. 4. Feature fusion: vote and weight

First, the three multi-objective optimization algorithms (NSGA-II, MOPSO, and MODE) perform parallel computations, completing genetic operations such as crossover and mutation, to obtain Pareto front feature subset collections. Each feature subset represents a chromosome, with binary encoding. Each gene position corresponds to a feature in the sequence. Subsequently, the Pareto front feature subsets from these algorithms are subjected to feature fusion. The voting method ranks features based on the number of votes each gene receives, ordering them from highest to lowest; features with zero votes are removed, resulting in the fused ordered feature subset. The weighting method further uses the number of votes as weights for the corresponding features, producing a weighted ordered feature subset. Finally, machine learning methods use the feature subsets obtained from feature fusion for model training.

## IV. EXPERIMENTAL DESIGN AND RESULT ANALYSIS

This section presents the experimental setup, including the datasets used, baseline methods, and research questions. For the experiments, the sampling techniques and multi-objective optimization algorithms discussed in this paper were implemented using Python. The hardware environment for the experiments consisted of an Intel(R) Core(TM) i5-9500 CPU @3.00GHz.

A. *Experimental Datasets*

The experiments utilized two well-known public software defect datasets: NASA and PROMISE [21]. NASA includes statistical information on several software projects and is a widely used datasets. Eight datasets were selected from NASA: CM1, KC3, MC1, MC2, MW1, PC2, PC4, and PC5. Table I provides basic statistical information about those datasets.

PROMISE includes data from the Eclipse project, Jedit software datasets, and other open-source datasets. Six datasets from PROMISE were selected: Ant-1.7, Camel-1.6, Jedit-4.3, Synapse-1.1, Poi-2.0, and Log4j-1.0. Table II provides basic statistical information about those datasets.

TABLE I. BASIC INFORMATION OF NASA.

| Dataset | Number of Modules | Number of Features | Number of Defective Modules | Number of Non-defective Modules | Defect Sample Rate |
|---|---|---|---|---|---|
| CM1 | 327 | 37 | 42 | 285 | 12.84% |
| KC3 | 200 | 39 | 36 | 164 | 18.00% |
| MC1 | 1988 | 38 | 46 | 1942 | 2.31% |
| MC2 | 125 | 39 | 44 | 81 | 35.20% |
| MW1 | 264 | 37 | 27 | 237 | 10.23% |
| PC2 | 1585 | 36 | 16 | 1569 | 1.01% |
| PC4 | 1287 | 37 | 177 | 1110 | 13.75% |
| PC5 | 1711 | 38 | 471 | 1240 | 27.53% |

TABLE II. BASIC INFORMATION OF PROMISE.

| Dataset | Number of Modules | Number of Features | Number of Defective Modules | Number of Non-defective Modules | Defect Sample Rate |
|---|---|---|---|---|---|
| Ant-1.7 | 745 | 20 | 166 | 579 | 22.28% |
| Camel-1.6 | 965 | 20 | 188 | 777 | 19.48% |
| Jedit-4.3 | 492 | 20 | 11 | 481 | 2.24% |
| Synapse-1.1 | 157 | 20 | 16 | 141 | 10.19% |
| Poi-2.0 | 314 | 20 | 37 | 277 | 11.78% |
| Log4j-1.0 | 135 | 20 | 34 | 101 | 25.19% |

B. *Handling Imbalanced Data*

The software defect prediction datasets contain many outlier samples, which can affect model training and reduce classification effectiveness. This study employs normalization by linearly transforming the original data. First, the *maxA* and *minA* of the attribute *A* are determined. Then, the original value of attribute *A* is mapped using the formula:

$$x' = \frac{x - minA}{maxA - minA} \quad (9)$$

To address the class imbalance issue in software defect datasets, this study uses Borderline SMOTE and Tomek Links algorithms. The experiments compare the effectiveness of these two sampling algorithms using evaluation metrics such as AUC, F-score, and Accuracy (ACC). To determine whether there are significant differences between the algorithms, Friedman and Wilcoxon Signed Rank tests are employed for statistical comparison of the experimental results.

Tables III and IV present the module statistics for the NASA and PROMISE datasets, highlighting a significant class imbalance in the original samples, with some datasets containing relatively few defective samples. To address this, the oversampling Borderline SMOTE algorithm and the hybrid SMOTE-Tomek algorithm were applied, effectively balancing the number of defective and non-defective samples.

TABLE III. STATISTICS OF NASA DATASET MODULES.

| Dataset | Original Samples | | Borderline-SMOTE | | SMOTE-Tomek | |
|---|---|---|---|---|---|---|
| | Non-defective | Defective | Non-defective | Defective | Non-defective | Defective |
| CM1 | 285 | 42 | 285 | 285 | 284 | 284 |
| KC3 | 164 | 36 | 164 | 164 | 161 | 161 |
| MC1 | 1942 | 46 | 1942 | 1942 | 1941 | 1941 |
| MC2 | 81 | 44 | 81 | 81 | 78 | 78 |
| MW1 | 237 | 27 | 237 | 237 | 235 | 235 |
| PC2 | 1569 | 16 | 1569 | 1569 | 1569 | 1569 |
| PC4 | 1110 | 177 | 1110 | 1110 | 1105 | 1105 |
| PC5 | 1240 | 471 | 1240 | 1240 | 1198 | 1198 |

To verify the effectiveness of the data sampling techniques on the model, experiments compared the Borderline SMOTE and SMOTE-Tomek algorithms. In Tables V to XVIII, comparisons were made between the original dataset, normalized dataset, normalized dataset using Borderline SMOTE, and normalized dataset using SMOTE-Tomek. The comparison metrics included AUC, F-score, and ACC.

TABLE IV. STATISTICS OF PROMISE DATASET MODULES.

| Dataset | Original Samples | | Borderline-SMOTE | | SMOTE-Tomek | |
|---|---|---|---|---|---|---|
| | Non-defective | Defective | Non-defective | Defective | Non-defective | Defective |
| Ant-1.7 | 579 | 166 | 579 | 579 | 567 | 567 |
| Camel-1.6 | 777 | 188 | 777 | 777 | 769 | 769 |
| Jedit-4.3 | 481 | 11 | 481 | 481 | 481 | 481 |
| Synapse-1.1 | 162 | 60 | 162 | 162 | 157 | 157 |
| Poi-2.0 | 277 | 37 | 277 | 277 | 272 | 272 |
| Log4j-1.0 | 101 | 34 | 101 | 101 | 101 | 101 |

TABLE V. NASA DATABASE: CM1.

| Multi-objective Optimization Algorithm | Evaluation Metric | Original Data | Normalized Data | Borderline SMOTE | SMOTE-Tomek |
|---|---|---|---|---|---|
| NSGA-II | AUC | 0.703 | 0.742 | 0.914 | **0.925** |
| | F-score | 0.000 | 0.000 | **0.859** | **0.859** |
| | ACC | **0.872** | **0.872** | 0.852 | 0.852 |
| MOPSO | AUC | 0.741 | 0.788 | **0.928** | **0.928** |
| | F-score | 0.080 | 0.000 | **0.871** | 0.866 |
| | ACC | **0.878** | 0.872 | 0.863 | 0.854 |
| MODE | AUC | 0.734 | 0.800 | 0.921 | **0.924** |
| | F-score | 0.080 | 0.080 | 0.868 | **0.877** |
| | ACC | **0.877** | **0.877** | 0.857 | 0.866 |

TABLE VI. NASA DATABASE: KC3.

| Multi-objective Optimization Algorithm | Evaluation Metric | Original Data | Normalized Data | Borderline SMOTE | SMOTE-Tomek |
|---|---|---|---|---|---|
| NSGA-II | AUC | 0.764 | 0.779 | 0.907 | **0.931** |
| | F-score | 0.275 | 0.275 | 0.849 | **0.858** |
| | ACC | 0.820 | 0.825 | 0.841 | **0.851** |
| MOPSO | AUC | 0.797 | 0.815 | 0.921 | **0.938** |
| | F-score | 0.275 | 0.329 | **0.856** | 0.842 |
| | ACC | 0.810 | 0.835 | **0.844** | 0.832 |
| MODE | AUC | 0.764 | 0.803 | 0.907 | **0.919** |
| | F-score | 0.120 | 0.329 | 0.833 | **0.836** |
| | ACC | 0.820 | **0.830** | 0.820 | 0.829 |

TABLE VII. NASA DATABASE: MC1.

| Multi-objective Optimization Algorithm | Evaluation Metric | Original Data | Normalized Data | Borderline SMOTE | SMOTE-Tomek |
|---|---|---|---|---|---|
| NSGA-II | AUC | 0.733 | 0.765 | **0.965** | 0.944 |
| | F-score | 0.000 | 0.000 | **0.935** | 0.869 |
| | ACC | **0.977** | **0.977** | 0.933 | 0.873 |
| MOPSO | AUC | 0.810 | 0.850 | **0.968** | 0.951 |
| | F-score | 0.000 | 0.000 | **0.934** | 0.864 |
| | ACC | **0.977** | **0.977** | 0.932 | 0.869 |
| MODE | AUC | 0.736 | 0.822 | **0.965** | 0.948 |
| | F-score | 0.00 | 0.000 | **0.895** | 0.875 |
| | ACC | **0.977** | **0.977** | 0.897 | 0.879 |

TABLE VIII. NASA DATABASE: MC2.

| Multi-objective Optimization Algorithm | Evaluation Metric | Original Data | Normalized Data | Borderline SMOTE | SMOTE-Tomek |
|---|---|---|---|---|---|
| NSGA-II | AUC | 0.744 | 0.836 | 0.864 | **0.887** |
| | F-score | 0.316 | 0.567 | **0.750** | 0.747 |
| | ACC | 0.704 | 0.745 | **0.754** | 0.745 |
| MOPSO | AUC | 0.781 | 0.840 | 0.881 | **0.907** |
| | F-score | 0.328 | 0.519 | 0.783 | **0.797** |
| | ACC | 0.704 | 0.746 | 0.771 | **0.796** |
| MODE | AUC | 0.768 | 0.815 | 0.866 | **0.891** |
| | F-score | 0.399 | 0.490 | 0.750 | **0.761** |
| | ACC | 0.713 | 0.720 | 0.754 | **0.758** |

TABLE IX. NASA DATABASE: MW1.

| Multi-objective Optimization Algorithm | Evaluation Metric | Original Data | Normalized Data | Borderline SMOTE | SMOTE-Tomek |
|---|---|---|---|---|---|
| NSGA-II | AUC | 0.812 | 0.805 | **0.941** | 0.914 |
| | F-score | 0.020 | 0.187 | **0.849** | 0.832 |
| | ACC | **0.894** | **0.894** | 0.848 | 0.830 |
| MOPSO | AUC | 0.831 | 0.836 | **0.946** | 0.940 |
| | F-score | 0.092 | 0.275 | 0.858 | **0.875** |
| | ACC | 0.894 | 0.902 | 0.859 | **0.866** |
| MODE | AUC | 0.828 | 0.836 | **0.943** | 0.929 |
| | F-score | 0.022 | 0.275 | **0.858** | 0.852 |
| | ACC | 0.879 | **0.902** | 0.859 | 0.845 |

TABLE X. NASA DATABASE: PC2.

| Multi-objective Optimization Algorithm | Evaluation Metric | Original Data | Normalized Data | Borderline SMOTE | SMOTE-Tomek |
|---|---|---|---|---|---|
| NSGA-II | AUC | 0.818 | 0.865 | **0.996** | 0.989 |
| | F-score | 0.000 | 0.000 | **0.984** | 0.969 |
| | ACC | 0.989 | 0.989 | **0.984** | 0.967 |
| MOPSO | AUC | 0.914 | 0.942 | **0.996** | 0.992 |
| | F-score | 0.000 | 0.000 | **0.980** | 0.968 |
| | ACC | **0.989** | **0.989** | 0.980 | 0.967 |
| MODE | AUC | 0.887 | 0.892 | **0.997** | 0.989 |
| | F-score | 0.000 | 0.000 | **0.983** | 0.969 |
| | ACC | **0.989** | **0.989** | 0.983 | 0.967 |

TABLE XI. NASA DATABASE: PC4.

| Multi-objective Optimization Algorithm | Evaluation Metric | Original Data | Normalized Data | Borderline SMOTE | SMOTE-Tomek |
|---|---|---|---|---|---|
| NSGA-II | AUC | 0.910 | 0.907 | **0.947** | 0.943 |
| | F-score | 0.361 | 0.365 | **0.901** | 0.885 |
| | ACC | 0.889 | 0.887 | **0.896** | 0.881 |
| MOPSO | AUC | 0.911 | 0.907 | 0.945 | **0.949** |
| | F-score | 0.332 | 0.415 | **0.897** | 0.896 |
| | ACC | 0.886 | **0.897** | 0.893 | 0.892 |
| MODE | AUC | 0.907 | 0.908 | **0.945** | **0.945** |
| | F-score | 0.359 | 0.326 | **0.896** | 0.895 |
| | ACC | 0.884 | 0.886 | 0.891 | **0.892** |

TABLE XII. NASA DATABASE: PC5.

| Multi-objective Optimization Algorithm | Evaluation Metric | Original Data | Normalized Data | Borderline SMOTE | SMOTE-Tomek |
|---|---|---|---|---|---|
| NSGA-II | AUC | 0.710 | 0.732 | 0.771 | **0.788** |
| | F-score | 0.265 | 0.206 | 0.670 | **0.725** |
| | ACC | **0.741** | **0.741** | 0.680 | 0.711 |
| MOPSO | AUC | 0.714 | 0.749 | 0.776 | **0.795** |
| | F-score | 0.265 | 0.199 | 0.728 | **0.740** |
| | ACC | 0.741 | **0.742** | 0.703 | 0.727 |
| MODE | AUC | 0.708 | 0.745 | 0.771 | **0.789** |
| | F-score | 0.072 | 0.265 | 0.671 | **0.724** |
| | ACC | 0.722 | **0.746** | 0.680 | 0.706 |

TABLE XIII. PROMISE DATABASE: ANT-1.7.

| Multi-objective Optimization Algorithm | Evaluation Metric | Original Data | Normalized Data | Borderline SMOTE | SMOTE-Tomek |
|---|---|---|---|---|---|
| NSGA-II | AUC | 0.830 | 0.815 | 0.860 | **0.868** |
| | F-score | 0.359 | 0.491 | **0.798** | 0.789 |
| | ACC | 0.813 | **0.819** | 0.793 | 0.797 |
| MOPSO | AUC | 0.828 | 0.821 | 0.867 | **0.878** |
| | F-score | 0.490 | 0.490 | **0.805** | 0.802 |
| | ACC | 0.815 | **0.817** | 0.799 | 0.800 |
| MODE | AUC | 0.828 | 0.816 | 0.867 | **0.875** |
| | F-score | 0.490 | 0.490 | **0.805** | 0.800 |
| | ACC | 0.815 | **0.817** | 0.799 | 0.799 |

TABLE XIV. PROMISE DATABASE: CAMEL-1.6.

| Multi-objective Optimization Algorithm | Evaluation Metric | Original Data | Normalized Data | Borderline SMOTE | SMOTE-Tomek |
|---|---|---|---|---|---|
| NSGA-II | AUC | 0.679 | 0.728 | **0.796** | 0.793 |
| | F-score | 0.029 | 0.057 | **0.747** | 0.730 |
| | ACC | 0.802 | **0.805** | 0.737 | 0.722 |
| MOPSO | AUC | 0.683 | 0.740 | **0.800** | 0.797 |
| | F-score | 0.056 | 0.057 | **0.752** | 0.745 |
| | ACC | **0.805** | **0.805** | 0.735 | 0.725 |
| MODE | AUC | 0.676 | 0.723 | **0.799** | 0.793 |
| | F-score | 0.056 | 0.057 | **0.752** | 0.746 |
| | ACC | 0.802 | **0.805** | 0.737 | 0.719 |

TABLE XV. PROMISE DATABASE: JEDIT-4.3.

| Multi-objective Optimization Algorithm | Evaluation Metric | Original Data | Normalized Data | Borderline SMOTE | SMOTE-Tomek |
|---|---|---|---|---|---|
| NSGA-II | AUC | 0.789 | 0.897 | **0.982** | 0.980 |
| | F-score | 0.000 | 0.000 | 0.924 | **0.925** |
| | ACC | **0.978** | **0.978** | 0.920 | 0.920 |
| MOPSO | AUC | 0.933 | 0.921 | **0.983** | 0.980 |
| | F-score | 0.000 | 0.000 | **0.929** | 0.926 |
| | ACC | **0.978** | **0.978** | 0.925 | 0.921 |
| MODE | AUC | 0.853 | 0.924 | **0.980** | 0.978 |
| | F-score | 0.000 | 0.000 | 0.922 | **0.923** |
| | ACC | **0.978** | **0.978** | 0.918 | 0.919 |

TABLE XVI. PROMISE DATABASE: SYNAPSE-1.1.

| Multi-objective Optimization Algorithm | Evaluation Metric | Original Data | Normalized Data | Borderline SMOTE | SMOTE-Tomek |
|---|---|---|---|---|---|
| NSGA-II | AUC | 0.807 | 0.831 | 0.833 | **0.866** |
| | F-score | 0.350 | 0.343 | 0.754 | **0.764** |
| | ACC | 0.770 | **0.775** | 0.757 | 0.761 |
| MOPSO | AUC | 0.816 | 0.838 | 0.839 | **0.874** |
| | F-score | 0.321 | 0.484 | 0.762 | **0.797** |
| | ACC | 0.765 | **0.802** | 0.763 | 0.784 |
| MODE | AUC | 0.814 | 0.839 | 0.832 | **0.873** |
| | F-score | 0.319 | 0.495 | 0.713 | **0.790** |
| | ACC | 0.766 | **0.802** | 0.729 | 0.780 |

TABLE XVII. PROMISE DATABASE: POI-2.0.

| Multi-objective Optimization Algorithm | Evaluation Metric | Original Data | Normalized Data | Borderline SMOTE | SMOTE-Tomek |
|---|---|---|---|---|---|
| NSGA-II | AUC | 0.812 | 0.774 | **0.928** | 0.896 |
| | F-score | 0.000 | 0.000 | **0.845** | 0.829 |
| | ACC | 0.882 | 0.882 | **0.837** | 0.814 |
| MOPSO | AUC | 0.812 | 0.803 | **0.931** | 0.898 |
| | F-score | 0.120 | 0.040 | **0.853** | 0.834 |
| | ACC | 0.892 | 0.885 | **0.854** | 0.821 |
| MODE | AUC | 0.795 | 0.786 | **0.928** | 0.898 |
| | F-score | 0.120 | 0.040 | **0.845** | 0.827 |
| | ACC | 0.892 | 0.885 | **0.838** | 0.814 |

After the software defects dataset was normalized, the AUC, F-score, and ACC of the models in the NASA dataset were all improved except for MW1 and PC4. In the PROMISE dataset, AUC, F-score and ACC were improved for all models except Camel-1.6 and Poi-2.0. CM1, MC1 and PC2 in NASA in the above dataset without sampling algorithm processing have generally higher model classification accuracy ACC, but it does not indicate a better model classification because its F-score value is zero. Meanwhile, the experimental results also demonstrated that the class imbalance problem would have a more serious effect on the model.

The experimental results show that the Borderline SMOTE algorithm performs relatively better on MW1, MC1, PC2, PC4, Camel-1.6, Jedite-4.3, Poi-2.0, and Log4j-1.0. The SMOTE-Tomek algorithm performs relatively better on CM1, KC3, MC2, PC5, Ant-1.7, and Synapse-1.1 datasets with relatively better performance.

TABLE XVIII. PROMISE DATABASE: LOG4J-1.0.

| Multi-objective Optimization Algorithm | Evaluation Metric | Original Data | Normalized Data | Borderline SMOTE | SMOTE-Tomek |
|---|---|---|---|---|---|
| NSGA-II | AUC | 0.854 | 0.880 | **0.930** | 0.912 |
| | F-score | 0.528 | 0.502 | **0.867** | 0.782 |
| | ACC | 0.823 | 0.815 | **0.852** | 0.797 |
| MOPSO | AUC | 0.872 | 0.886 | **0.934** | 0.921 |
| | F-score | 0.554 | 0.530 | **0.862** | 0.810 |
| | ACC | 0.829 | 0.823 | **0.862** | 0.817 |
| MODE | AUC | 0.854 | 0.880 | **0.925** | 0.918 |
| | F-score | 0.528 | 0.502 | **0.835** | 0.808 |
| | ACC | 0.823 | 0.815 | **0.832** | 0.816 |

The Wilcoxon Signed Rank Test was used to compare the Borderline SMOTE and SMOTE-Tomek algorithms. Table XIX and table XX shows whether there were significant differences between the sampling algorithms, with AUC as the evaluation metric. The null hypothesis H0 posits that the evaluation value distributions of each sample set are the same under both algorithms, while the alternative hypothesis H1 suggests they are different. If the p-value is greater than 0.05, the distributions are considered the same; if less than 0.05, they are considered different.

TABLE XIX. DIFFERENCES IN SAMPLING ALGORITHMS ON NASA.

| Algorithm | statistic | p-value | Significant Difference |
|---|---|---|---|
| NSGA-II | 16.000 | 0.844 | no |
| MOPSO | 9.500 | 0.446 | no |
| MODE | 11.000 | 0.612 | no |

TABLE XX. DIFFERENCES IN SAMPLING ALGORITHMS ON PROMISE.

| Algorithm | statistic | p-value | Significant Difference |
|---|---|---|---|
| NSGA-II | 9.000 | 0.844 | no |
| MOPSO | 9.000 | 0.844 | no |
| MODE | 10.000 | 1.000 | no |

The test results indicate that the null hypothesis H0 is accepted, meaning that the two sampling algorithms perform similarly in terms of model improvement on the NASA and PROMISE datasets. The p-values for both the Borderline SMOTE algorithm and the SMOTE-Tomek algorithm are greater than 0.05, indicating that there is no significant difference between the two.

*C. Feature Selection Based on Multi-objective Optimization*

The machine learning method used for classifying software defect prediction samples is Support Vector Machine (SVM), implemented using the sklearn.svm.SVC package in Python with the radial basis function (RBF) kernel (kernel='rbf'). The balanced datasets were divided into training and testing sets using 10-fold cross-validation (K=10). This setup ensures reliable evaluation performance without excessive computational overhead and effectively tests the model's generalization ability. Feature selection was addressed using several multi-objective optimization algorithms, which aimed to minimize the number of features while maximizing the model's performance (AUC). The parameters for these algorithms are provided in Tables XXI, XXII, and XXIII.

TABLE XXI. NSGA-II ALGORITHM PARAMETERS.

| Parameter Name | Value |
| --- | --- |
| Population Size (nPop) | 100 |
| Maximum Iteration (nIter) | 100 |
| Crossover Probability (Pc) | 0.6 |
| Mutation Probability (Pm) | 0.1 |
| Crossover Distribution Index (etaC) | 1 |
| Mutation Distribution Index (etaM) | 1 |

TABLE XXII. MOPSO ALGORITHM PARAMETERS.

| Parameter Name | Value |
| --- | --- |
| Population Size (nPop) | 100 |
| Maximum Iterations (nIter) | 100 |
| Particle Size (nChr) | The chromosome length |
| Maximum Archive Size (nAr) | 100 |
| Velocity Update Parameters (c1 and c2) | 1.49 and 2 |
| Maximum Velocity (Vmax) | 1 |
| Minimum Velocity (Vmin) | -1 |
| Number of Grid Divisions (M*M) | 50 |

TABLE XXIII. MODE ALGORITHM PARAMETERS.

| Parameter Name | Value |
| --- | --- |
| Population Size (nPop) | 100 |
| Maximum Iterations (nIter) | 100 |
| Crossover Probability (Pc) | 0.5 |
| Scaling Factor (F) | 0.5 |

To compare the performance of NSGA-II, MOPSO, and MODE in feature selection for software defect prediction, this section presents the Pareto fronts and solution space distributions obtained through experiments as shown in Fig. 5 and Fig. 6. Blue dots represent the solution space, recording the feature subsets searched by the algorithms. Red dots indicate the Pareto front, with each red dot representing a non-dominated feature subset. For comparative purposes, each row in the figure represents the same dataset, with three columns for the three multi-objective optimization algorithms.

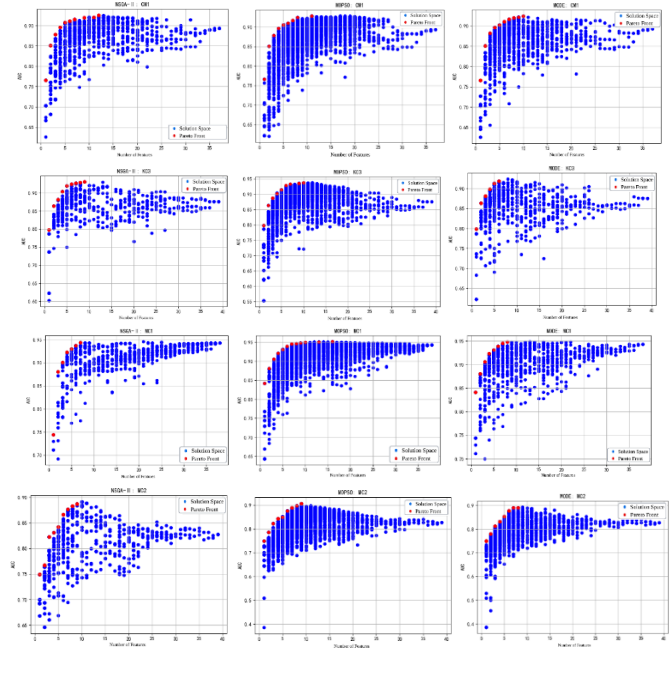

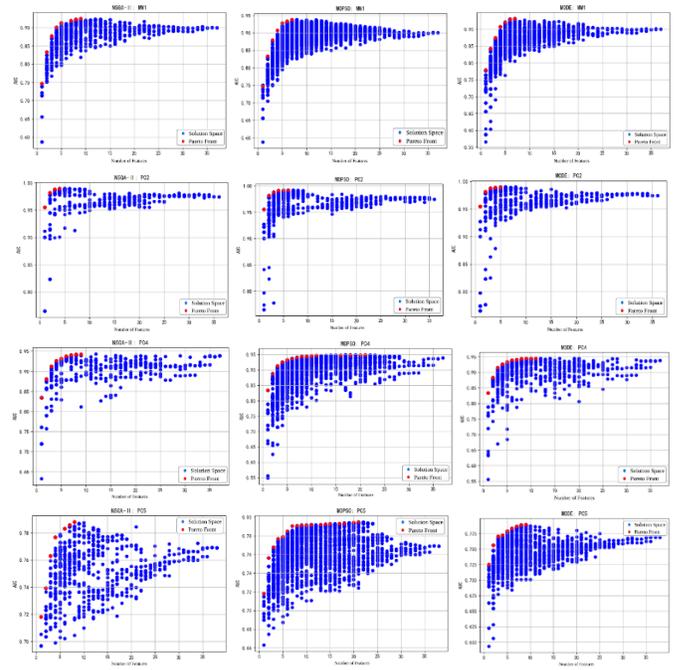

Fig. 5. NASA database: Pareto front of multi-objective optimization

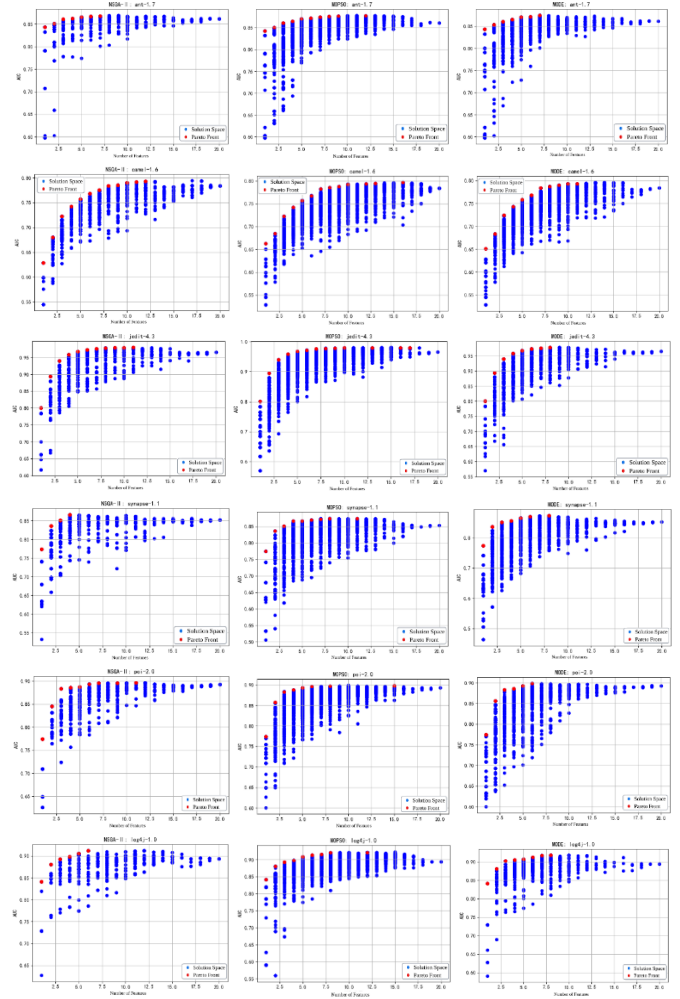

Fig. 6. PROMISE database: Pareto front of multi-objective optimization

The experimental results show:

- On NASA datasets: MOPSO's solution space range is larger, indicating higher population diversity. The MODE algorithm follows, with the NSGA-II algorithm having the smallest solution space range. The scatter points forming the Pareto front in each dataset show that the model performance evaluation metric (AUC) begins to stabilize when the number of features reaches seven or more. As the number of features increases, AUC generally decreases, indicating the presence of irrelevant and redundant features in the software defect datasets, validating the high-dimensional feature issue.

- On PROMISE datasets: the MOPSO and MODE algorithms have similar solution space ranges, indicating comparable population diversity. The NSGA-II algorithm has a relatively smaller solution space range. Like the NASA datasets, the scatter points forming the Pareto front show that AUC decreases when the number of features exceeds 15, indicating the presence of irrelevant and redundant features.

Table XXIV provide the optimal AUC values for the Pareto fronts of the NSGA-II, MOPSO, and MODE algorithms, with "BS" indicating the use of the Borderline SMOTE oversampling algorithm and "ST" indicating the use of the SMOTE-Tomek hybrid sampling algorithm. Bold values indicate the maximum values for each dataset, with the second-highest AUC values underlined.

TABLE XXIV. OPTIMAL AUC VALUES FOR MULTI-OBJECTIVE OPTIMIZATION ALGORITHMS.

| Dataset | NSGA-II | | MOPSO | | MODE | |
|---|---|---|---|---|---|---|
| | BS | ST | BS | ST | BS | ST |
| CM1 | 0.914 | <u>0.925</u> | **0.928** | **0.928** | 0.921 | 0.924 |
| KC3 | 0.907 | <u>0.931</u> | 0.921 | **0.938** | 0.907 | 0.919 |
| MC1 | <u>0.965</u> | 0.944 | **0.968** | 0.951 | <u>0.965</u> | 0.948 |
| MC2 | 0.864 | 0.887 | 0.881 | **0.907** | 0.866 | <u>0.891</u> |
| MW1 | 0.941 | 0.914 | **0.946** | 0.940 | <u>0.943</u> | 0.929 |
| PC2 | <u>0.996</u> | 0.989 | <u>0.996</u> | 0.992 | **0.997** | 0.989 |
| PC4 | <u>0.947</u> | 0.943 | 0.945 | **0.949** | 0.945 | 0.945 |
| PC5 | 0.771 | 0.788 | 0.776 | **0.795** | 0.771 | <u>0.789</u> |
| Ant-1.7 | 0.860 | 0.868 | 0.867 | **0.878** | 0.867 | <u>0.875</u> |
| Camel-1.6 | 0.796 | 0.793 | **0.800** | 0.797 | <u>0.799</u> | 0.793 |
| Jedit-4.3 | <u>0.982</u> | 0.980 | **0.983** | 0.980 | 0.980 | 0.978 |
| Synapse-1.1 | 0.833 | 0.866 | 0.839 | **0.874** | 0.832 | <u>0.873</u> |
| Poi-2.0 | <u>0.928</u> | 0.896 | **0.931** | 0.898 | <u>0.928</u> | 0.898 |
| Log4j-1.0 | <u>0.930</u> | 0.912 | **0.934** | 0.921 | 0.925 | 0.918 |

Fig. 7 compares the differences between the algorithms, where the blue dashed line represents the average, the red solid line represents the median, and the red dots indicate outliers. In the left figure, the optimal algorithm is MOPSO_ST, with an average of 0.925 and a median of 0.937. In the right figure, the optimal algorithm is MOPSO_BS, with an average of 0.892 and a median of 0.899.

In NASA, the medians of all algorithms are greater than the average, indicating negative skewness. The MOPSO_ST algorithm has the smallest box height, indicating minimal variation across different datasets and the highest stability, with the highest average and median values. The BS algorithm typically has slightly higher box heights than the ST algorithm, indicating relatively lower model stability. The outliers in the figure are from the PC5 dataset, but an outlier in the MOPSO_ST algorithm appears in the Jedit-4.3 dataset. The median values of the six algorithms are all greater than 0.926, and the average values are all greater than 0.914.

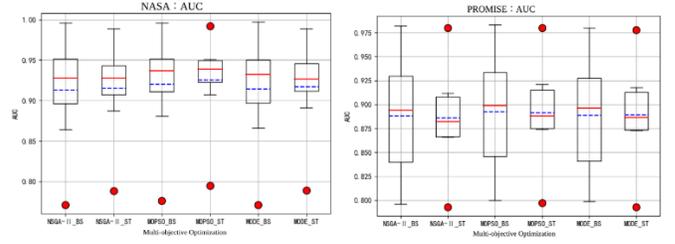

Fig. 7. Box plot of AUC values for multi-objective optimization algorithms

In PROMISE, the median values of the NSGA-II_BS, MOPSO_BS, and MODE_BS algorithms are greater than the average values, indicating negative skewness (or left skewness). In contrast, the NSGA-II_ST, MOPSO_ST, and MODE_ST algorithms have average values greater than the median values, indicating positive skewness (or right skewness). The MOPSO_ST algorithm has the highest median and average values, but a larger box height indicates relatively lower model stability across different datasets. The box height variation between sampling algorithms is evident, indicating that the choice of sampling algorithm significantly affects model applicability. The BS algorithm typically exhibits greater fluctuation in model evaluation values, while the ST algorithm is generally more stable. The outliers are from the Camel-1.6 and Jedit-4.3 datasets. The median values of the six algorithms are all greater than 0.882, and the average values are all greater than 0.885.

TABLE XXV. DIFFERENCES IN MULTI-OBJECTIVE OPTIMIZATION ALGORITHMS ON NASA.

| Algorithm | statistic | p-value | Significant Difference |
|---|---|---|---|
| Borderline SMOTE | 6.222 | 0.045 | yes |
| SMOTE-Tomek | 12.968 | 0.002 | yes |

TABLE XXVI. DIFFERENCES IN MULTI-OBJECTIVE OPTIMIZATION ALGORITHMS ON PROMISE.

| Algorithm | statistic | p-value | Significant Difference |
|---|---|---|---|
| Borderline SMOTE | 8.273 | 0.016 | yes |
| SMOTE-Tomek | 8.000 | 0.018 | yes |

Based on the experimental results of the multi-objective optimization algorithm AUC values presented in the previous section, Table XXV and table XXVI show the Friedman test results for the three multi-objective optimization algorithms to determine whether there are significant differences, with AUC as the evaluation metric. The null hypothesis H0 posits that the evaluation value distributions of each sample set are the same under the three algorithms, while the alternative hypothesis H1 suggests that the evaluation value distributions differ for at least one algorithm. If the p-value is greater than 0.05, the distributions are considered the same; if less than 0.05, they are considered different.

The above results indicate that the alternative hypothesis H1 is accepted, meaning there is a significant difference in the model evaluation metrics for at least one of the NSGA-II, MOPSO, and MODE. The Friedman test chart visually displays the test results as shown in Fig. 8. The y-axis represents the three algorithms, while the x-axis represents the mean ranks. Each algorithm's mean rank is displayed with a dot, and the horizontal lines centered on the dots indicate the critical difference (CD) range. If the horizontal lines overlap between two algorithms, it means there is no significant difference between them; otherwise, there is a significant difference. The test results indicate that:

- On the NASA datasets, when using the Borderline SMOTE (BS) algorithm, no significant performance differences are observed among the NSGA-II, MOPSO, and MODE algorithms. However, with the SMOTE-Tomek (ST) algorithm, significant differences arise: MOPSO outperforms both NSGA-II and MODE, although there is no significant difference between NSGA-II and MODE.

- On the PROMISE datasets, the Borderline SMOTE (BS) algorithm shows MOPSO performing significantly better than MODE, while no significant difference is found between MOPSO and NSGA-II, or between MODE and NSGA-II. When the SMOTE-Tomek (ST) algorithm is applied, MOPSO again outperforms NSGA-II, but there are no significant differences between NSGA-II and MODE or between MOPSO and MODE.

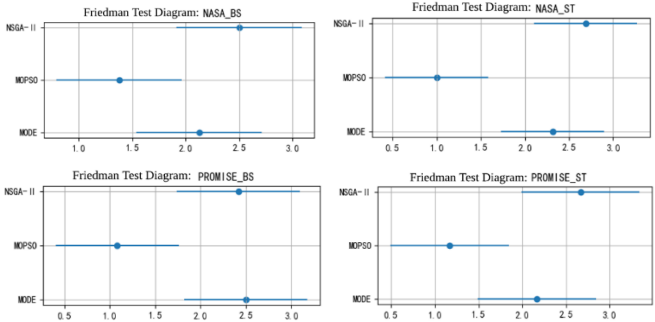

Fig. 8. Friedman test diagram

### D. Feature Fusion Based on Hybrid Sampling and Multi-objective Optimization

*1) Feature Fusion Based on Multi-objective Optimization Algorithms.*

This section evaluates the effectiveness of the proposed feature fusion methods (vote_ST, weight_ST) by comparing them with Greedy Forward Selection, Pearson Correlation Coefficient, and Fisher Discrimination Criterion [23], with the results presented in Fig. 9 and Fig. 10.

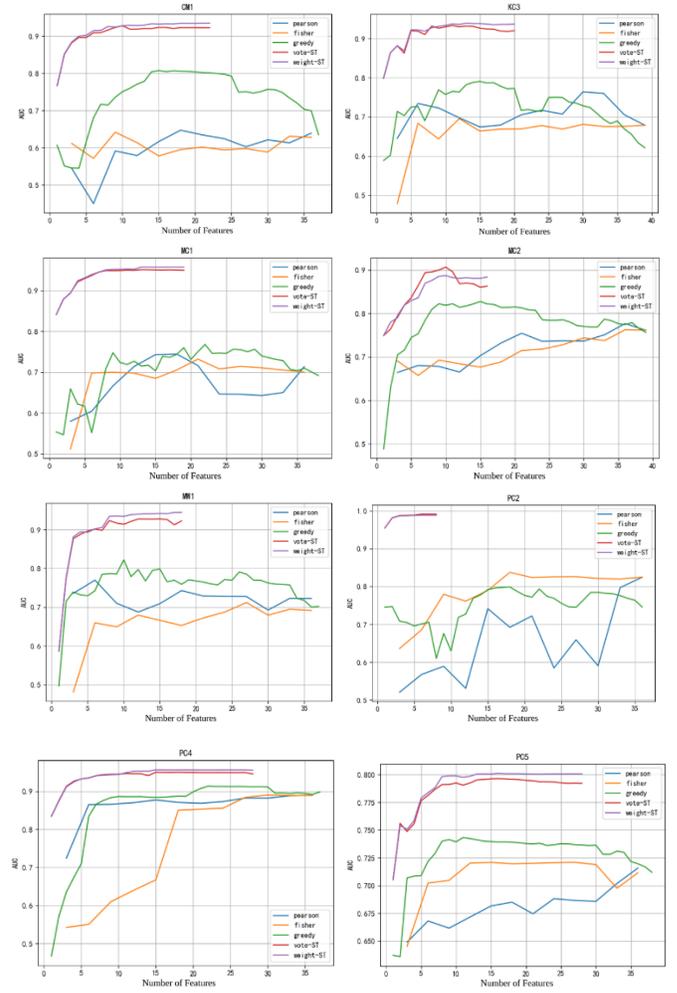

Fig. 9. Result on NASA datasets.

The experimental results indicate that feature fusion based on hybrid sampling and multi-objective optimization algorithms significantly improves AUC performance compared to Greedy Forward Selection, Pearson Correlation Coefficient, and Fisher Discrimination Criterion. The feature fusion methods perform well on both NASA and PROMISE datasets.

*2) Voting and Weighting Feature Fusion Methods.*

Table XXVII records the optimal AUC values for the feature fusion methods (vote_BS, weight_BS, vote_ST and weight_ST) and Pearson, Fisher, and Greedy on the NASA and PROMISE datasets.

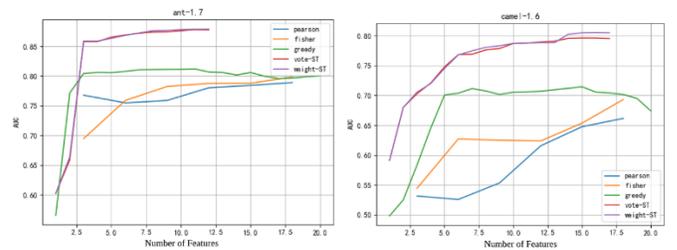

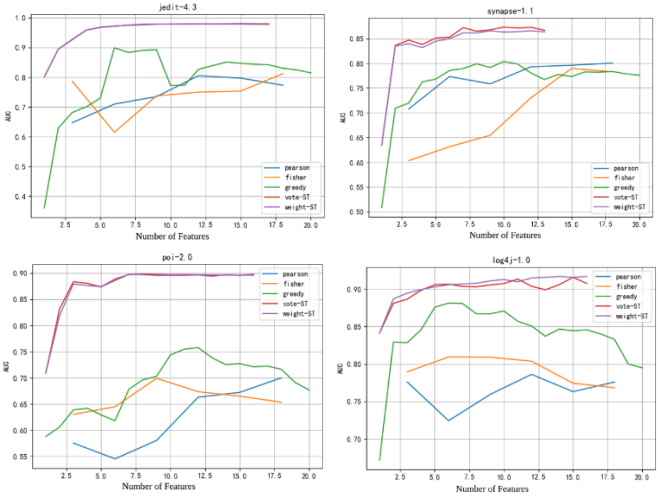

Fig. 10. Result on PROMISE datasets.

TABLE XXVII. OPTIMAL AUC VALUES FOR FEATURE FUSION METHODS.

| Dataset | pearson | fisher | greedy | vote_BS | weight_BS | vote_ST | weight_ST |
|---|---|---|---|---|---|---|---|
| CM1 | 0.647 | 0.641 | 0.807 | 0.929 | <u>0.933</u> | 0.927 | **0.934** |
| KC3 | 0.764 | 0.695 | 0.791 | 0.921 | 0.921 | <u>0.933</u> | **0.939** |
| MC1 | 0.744 | 0.732 | 0.768 | <u>0.969</u> | **0.978** | 0.951 | 0.958 |
| MC2 | 0.777 | 0.762 | 0.827 | 0.872 | 0.872 | **0.906** | <u>0.887</u> |
| MW1 | 0.770 | 0.712 | 0.822 | <u>0.946</u> | **0.951** | 0.926 | 0.939 |
| PC2 | 0.825 | 0.837 | 0.799 | **0.997** | <u>0.996</u> | 0.992 | 0.988 |
| PC4 | 0.889 | 0.890 | 0.913 | 0.948 | <u>0.953</u> | 0.950 | **0.956** |
| PC5 | 0.716 | 0.721 | 0.743 | 0.774 | 0.778 | <u>0.796</u> | **0.801** |
| Ant-1.7 | 0.789 | 0.798 | 0.812 | 0.867 | 0.866 | <u>0.878</u> | **0.879** |
| Camel-1.6 | 0.662 | 0.693 | 0.715 | <u>0.801</u> | 0.796 | 0.796 | **0.805** |
| Jedit-4.3 | 0.805 | 0.812 | 0.899 | <u>0.981</u> | **0.983** | 0.980 | 0.979 |
| Synapse-1.1 | 0.800 | 0.790 | 0.804 | 0.840 | 0.840 | **0.873** | <u>0.866</u> |
| Poi-2.0 | 0.700 | 0.699 | 0.758 | **0.929** | <u>0.928</u> | 0.898 | 0.899 |
| Log4j-1.0 | 0.786 | 0.809 | 0.881 | <u>0.931</u> | **0.940** | 0.916 | 0.917 |

The results vary across datasets for the four algorithms:

- For the NASA datasets, when the sampling algorithm is consistent, the weighting method generally achieves better AUC values than the voting method, as it considers feature weights. However, the voting method achieves relatively better AUC values on the PC2 dataset and some MC2 segments. When the feature fusion method is consistent, the hybrid sampling algorithm (SMOTE-Tomek) generally achieves better model evaluation metrics (AUC) than the oversampling algorithm (Borderline SMOTE). However, the oversampling algorithm performs better on the PC2 and MC1 datasets.

- For the PROMISE datasets, when the sampling algorithm is consistent, the model evaluation metrics (AUC) are generally similar for the voting and weighting methods, except for the Synapse-1.1 and Log4j-1.0 datasets. When the feature fusion method is consistent, the hybrid sampling algorithm performs relatively better on the Ant-1.7 and Synapse-1.1 datasets, while the oversampling algorithm achieves better AUC values on the other datasets.

This section also conducted Friedman and Nemenyi post hoc tests for the four algorithms. The null hypothesis H0 posits that the AUC distributions of the four algorithms are the same, while the alternative hypothesis H1 suggests that the AUC distributions differ for at least one algorithm. The p-values for the NASA and PROMISE datasets are 0.404 and 0.805, respectively, both greater than 0.05, indicating no significant difference among the four algorithms. The following figures show the Friedman test charts for the feature fusion methods (voting and weighting). The x-axis represents the mean ranks, and the y-axis represents the four algorithms. Fig. 11 show that the Weight_ST and Weight_BS algorithms have the highest mean ranks, while the voting methods have relatively lower mean ranks. However, Vote_BS has the highest mean rank in the PROMISE dataset.

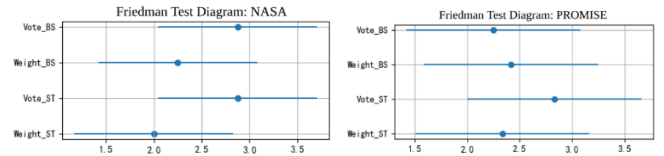

Fig. 11. Feature fusion: Friedman test plots for voting and weighting methods

### E. Threat to Validity

This study acknowledges several potential threats to validity. The datasets used, primarily from NASA and PROMISE, are widely recognized in the software engineering community, but they may not represent all types of software projects, particularly those from different domains or with unique characteristics. As a result, the generalizability of the findings to other datasets or real-world projects might be limited. Additionally, the choice of parameters for the multi-objective optimization algorithms (e.g., NSGA-II, MOPSO, MODE) was based on standard settings, which may not be optimal for all datasets, potentially influencing the results. The study employs AUC, F-score, and ACC as the primary evaluation metrics. While these metrics are standard in classification tasks, they may not fully capture all aspects of model performance, such as robustness to noise or the ability to handle extremely imbalanced data. Furthermore, statistical tests like the Friedman and Wilcoxon Signed Rank tests were used to compare the performance of different algorithms. However, the sample size and variability within the datasets could affect the power of these tests, potentially leading to Type I or Type II errors. Additionally, the study assumes that classifiers produce consistent and stable results across different datasets, which may not always hold true.

## V. CONCLUSION

This paper investigates issues related to software defect prediction, focusing on challenges such as class imbalance in data, high feature dimensionality, and low predictive accuracy. The study approaches these problems from the perspectives of data sampling and feature selection. It explores sampling techniques to address class imbalance and utilizes multi-objective optimization algorithms to tackle high-dimensional feature selection problems, establishing appropriate optimization goals. A software defect prediction model based on hybrid sampling and multi-objective optimization was developed. The study also examines feature fusion methods to balance the selection tendencies of feature subsets, addressing issues of model. The ultimate goal is to globally balance the dataset, remove redundant features, and improve prediction accuracy.

The experimental results demonstrate that the proposed feature fusion based on hybrid sampling and multi-objective optimization algorithms offers significant advantages. The feature fusion methods performed well on both the NASA and PROMISE datasets. In the NASA dataset, the optimal result was achieved with the Weight_BS method, with an average AUC value of 0.923 and a median of 0.942. In the PROMISE dataset, the Weight_BS method also had the highest average AUC value of 0.892, while the Vote_BS method had the highest median value of 0.898. Except for the MC2 and Synapse-1.1 datasets, where MOPSO_ST had relatively high AUC values, the feature fusion algorithms achieved relatively high AUC values across the remaining datasets.